\documentclass[conference]{./pvsctran}
\usepackage{graphicx}
\usepackage{hyperref}
\usepackage{graphicx}
\usepackage{amssymb}
\usepackage{amsmath}
\usepackage[super]{nth}
\usepackage{multicol}

\usepackage{cite}

\ifCLASSINFOpdf
\else
\fi
\usepackage[ruled,vlined,linesnumbered]{algorithm2e}
\usepackage{url}

\usepackage{siunitx}
\usepackage{url}

\title{\LARGE Signal Processing on PV Time-Series Data: \\ Robust Degradation Analysis without Physical Models}

\author{ \IEEEauthorblockN{\large Bennet~Meyers, Michael~Deceglie, Chris~Deline, and Dirk~Jordan} \\
\IEEEauthorblockA{\large Stanford University and SLAC National Accelerator Laboratory, Menlo Park, CA, USA \\ National Renewable Energy Laboratory, Golden CO, USA}}

\begin{document}
	
\setlength{\columnsep}{0.25in}
\maketitle

\begin{abstract}
	A novel unsupervised machine learning approach for analyzing time-series data is applied to the topic of photovoltaic (PV) system degradation rate estimation, sometimes referred to as energy-yield degradation analysis. This approach only requires a measured power signal as an input---no irradiance data, temperature data, or system configuration information. We present results on a data set that was previously analyzed and presented by NREL using \texttt{RdTools}, validating the accuracy of the new approach and showing increased robustness to data anomalies while reducing the data requirements to carry out the analysis.
\end{abstract}

\section{Introduction}
A large amount of research has been published on methods for estimating PV module and system degradation rates (for an overview, see~\cite{Phinikarides2014}). This work focuses on the estimation of system-level degradation from historical time-series power data. Recent advances in this area include the year-on-year (YOY) estimation method~\cite{Hasselbrink2013}, using clear sky models for robustness to irradiance sensor issues~\cite{Jordan2018}, and the development of \texttt{RdTools} to standardize and automate the process of estimating system degradation rates from historical time-series data~\cite{rdtools}.

Typically, the process for estimating the degradation rate of PV systems requires three steps: normalization, filtering, and data analysis. A standard implementation of this process is described in detail in~\cite{Jordan2018}. The normalization process requires the use of a physical model to estimate the expected power output of the system. The model can be based on a simple performance ratio calculation or a detailed DC performance model (such as the Sandia Array Performance Model~\cite{King2004}). Even with the standardization provided by \texttt{RdTools}, there are many decisions that must be made by an analyst attempting to implement this process: what performance model to use, what source of irradiance data to use (on-site measurements, satellite-based measurements, a clear sky model, etc.), how to estimate plane-of-array irradiance from available data, and how to estimate the cell temperature of the system, to name a few. In addition, all methods also require knowledge of the system configuration including some combination of site location, mounting configuration (tilt and azimuth, if fixed-tilt), and module technology. This general approach works very well for analyzing utility-scale PV systems, which tend to be well characterized, data rich, and of sufficient size to justify hand-cleaning of data and hand-tuning of physical models.

Unfortunately, many real-world data sets of distributed rooftop PV systems lack correlated irradiance and meteorological data and system configuration parameters. Additionally, analysis of distributed rooftop systems must be done at a scale that precludes any hand-cleaning or tuning on a site-by-site basis. In light of this, we present an alternative method for estimating the bulk degradation of PV systems that requires no additional information, data, or site models. Based on a method for estimating clear sky system power directly from power data (presented at WCPEC-7/PVSC-45~\cite{Meyers2018}), this novel approach utilizes recent results from the area of optimization and unsupervised machine learning, implementing a domain-specific form of \textit{generalized low rank models} (GLRM)~\cite{Udell2016}. We refer to the clear sky estimation methodology as Statistical Clear Sky Fitting (\texttt{SCSF}). Building on this approach, we develop a method for estimating YOY system degradation as part of the GLRM fitting procedure. We refer to the SCSF approach to data fitting and degradation estimation as ``model-less'' in the sense that a physical model of the PV system is not employed and to distinguish our approach from the standard process described in the previous paragraph. Strictly speaking, we are still fitting a mathematical model to observed data, but we use a statistical, time-series model.

This paper presents three topics: the methodology of including a degradation rate estimation in the \texttt{SCSF} fitting procedure, a discussion on algorithm tuning parameters and their impact on degradation estimation, and a fleet-scale application of \texttt{SCSF}. We  apply \texttt{SCSF} to a data set of over 500 PV systems in the United States, which was previously analyzed using \texttt{RdTools} \cite{Deceglie2018}. We present a comparison between the model-less \texttt{SCSF} approach and the \texttt{RdTools}-based approach. We find that the model-less approach agrees very well with \texttt{RdTools} on average over all sites, while having a lower variance around that average and being more resilient to common data errors. The algorithm is available as open-source software here: \url{https://github.com/slacgismo/StatisticalClearSky}.

\section{Methodology}

\subsection{Background on \text{SCSF}}

\begin{figure}[t]
	\centering
	\includegraphics[width=\columnwidth]{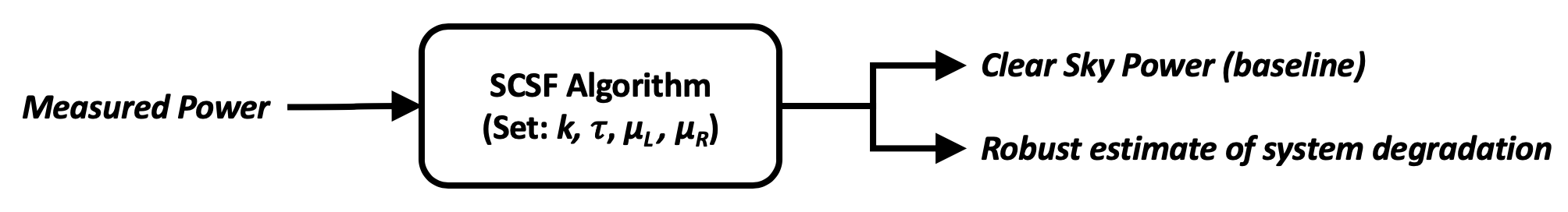}
	\caption{Functional diagram of the Statistical Clear Sky Fitting (SCSF) procedure. The tuning parameters are described in Section~\ref{ss:tune-param}.}
	\label{fig:function_diagram}
\end{figure}

For PV researchers and professionals, \texttt{SCSF} can be thought of as an abstract function that takes in measured power data and returns an estimate of the clear sky power output of the system, as shown in Figure~\ref{fig:function_diagram}. As described in Section~\ref{ss:deg-est}, the previously presented methodology has been extended to include estimation of YOY system degradation.

\texttt{SCSF} is comprised of two parts: (1) a mathematical model of PV power data over time and (2) an algorithm for optimally fitting this model to observed data~\cite{Meyers2018}. This approach exploits approximate periodicity in the PV power signal on daily and yearly time-scales to help separate the clear sky behavior of the system from all other dynamics. At a high level, the procedure involves forming the time-series data into a matrix and then finding a \textit{low-rank} approximation of that matrix~\cite{Udell2016}. Fitting the model requires solving a non-convex optimization problem (called the ``\texttt{SCSF} Problem''), and the algorithm presented in~\cite{Meyers2018} provides an effective heuristic that approximately solves the problem by solving a series of convex optimization problems, given a reasonable starting point.

\subsection{Related work}

The theory of \texttt{SCSF} relates to a few different domains from the fields of applied mathematics and photovoltaic system performance engineering. From a mathematical perspective, this work relates to both time-series analysis and low-rank matrix approximation. From a PV performance perspective, this work relates to clear sky system power modeling and system degradation analysis. Insights from these different fields provide the basis for developing \texttt{SCSF} and the approach to degradation analysis presented in this paper. 

\subsubsection{Time-series analysis}

Time-series data is a sequence of data points indexed in time order~\cite{Brillinger1981}, and this work relates to the subarea of time series analysis known as ``curve fitting,'' which involves both interpolation and smoothing of data with respect to time~\cite{Hastie2009}. The \texttt{SCSF} algorithm constructs an estimate of the clear sky power output, which smooths through the data on days that are approximately clear and interpolates through days that are cloudy. The algorithm implements a kind of nonparametric curve fitting that is conceptually similar to approaches such as weighted smoothing splines~\cite{Davies2008} and weighted local regression~\cite{Muller1987}. However, \texttt{SCSF} differs from these general approaches by employing bespoke techniques for analyzing PV power data. In essence, the algorithm estimates local statistics (percentile calculation) on an assumed \textit{cyclostationary} signal, i.e. a signal that has statistical properties that vary cyclically with time. This is accomplished by employing an asymmetric loss function and modeling time dependency of the data on three different time scales. These dependencies are (1) sequential times within a day should be smooth, (2) the same time on sequential days should be close to the same value, and (3) the same day on sequential years should have the same shape~\cite{Meyers2018}. In Section \ref{ss:deg-est}, we extend the model of the third time dependency to allow for a calculation of a YOY degradation rate, while maintaining the constraint that the shape of the signal on a daily scale has the same shape.

\subsubsection{Low-rank matrix approximation}

The \texttt{SCSF} algorithm solves a low-rank matrix approximation problem~\cite{Udell2016} over a matrix embedding of the observed data, as described in~\cite{Meyers2018}. Therefore, \text{SCSF} is closely related to other low-rank approximation techniques such as principle component analysis, invented by Karl Pearson in 1901~\cite{Pearson1901}, and singular value decomposition, formalized for rectangular matrices in 1936 by Eckart and Young~\cite{Eckart1936}. A modern treatement of the subject and its many generalizations is given in~\cite{Udell2016}. Recent advances in this subject include the ability to deal with missing values in the data matrix (``matrix completion''), which is included in the \texttt{SCSF} approach. The \texttt{SCSF} algorithm essentially solves a constrained dictionary learning problem~\cite{Tosic2011} on an ordered collection of signals, each representing one day of power output with many ``missing days'' representing cloudy days in the measured data set. The resulting factorization provides a set of vectors and a set of mixing factors. The vectors can be combined linearly to create any clear day within the measured data set, and the mixing factors describe how much of each vector to use on any given day. Through this separation of intra-day variability and inter-day variability in the observed signal, we are able to enforce constraints like smoothness and periodicity at sub-hourly, daily, and yearly timescales, which allows for interpolation through the cloudy periods in the measured data. 

\subsubsection{Clear sky models}

As a model of clear sky system behavior, \texttt{SCSF} is conceptually related to geometric sun position modeling~\cite{Reda2004, Reda2007} and atmospheric clear sky modeling~\cite{INEICHEN2008, INEICHEN2016}, especially recent work to ``tune'' these models to observed site data~\cite{Reno2016}. However, because \texttt{SCSF} is not constrained by physical models, it can easily model ``abnormal'' system behavior such as regular shading. Unlike all these cited appraoches, \texttt{SCSF} does not model the underlying geometry or physics; it is a completely data driven, non-parametric model that ``filters'' the observed data for the clear sky signal. In this way, the system is its own model, establishing the repeating, periodic nature of the signal over time. The resulting clear sky power estimate, represents the expected power output of the system on a given day of the year and year of operation in the absence of weather, soiling, or operational issues. The effects of soiling are filtered out by making year-over-year comparisons of energy output and by enforcing ``stiffness'' on the energy component within the clear sky estimate; it cannot fluctuate too rapidly. Notably, the effects of regular shading on the system would be included in the clear sky estimate.

\subsubsection{Degradation analysis}

As described in Section \ref{ss:deg-est}, degradation is estimated with \texttt{SCSF} as a year-over-year percent change in energy. Conceptually, the model of degradation rate is similar to the year-on-year approaches described in \cite{Hasselbrink2013, Jordan2018, Deceglie2018}. A comparison of YOY approaches to linear regression approaching for bulk system degradation rate estimation is given in~\cite{Curran2019}, while a review of other non-YOY approaches is given in~\cite{Phinikarides2014}. However, the \text{SCSF} estimation of degradation is fundamentally different than previous approaches. In previous work, the measured data are normalized by an imperfect model of expected system performance and then statistics---either linear regression or yean-over-year analysis---are applied to measure the long term degradation. This method is highly dependent on the model employed and the input data to that model. Recent work has suggested that using a clear sky model instead of performance model based on measured irradiance and temperature can lead to improved accuracy~\cite{Jordan2018}. The \text{SCSF} method takes this concept one step further, removing the need to have any physical model of the system or its location whatsoever. Degradation is simply interpreted as the change in the output of the optimal clear sky signal over time.

\subsection{Estimation of degradation}\label{ss:deg-est}

We expect the shape of the clear sky signal to repeat year-over-year, but we expect and do observe that the energy output reduces over time. In other words, a degradation signal is present in the power data. This means that fitting a clear sky model to multi-year data sets requires relaxing exact yearly periodicity so that the energy content changes over time. 

Estimating this degradation, however, is mathematically difficult due to the non-convex nature of YOY degradation. As stated previously, the \texttt{SCSF} Problem is non-convex; however, it has a particular structure known as \textit{biconvexity}~\cite{Gorski2007}. This structure is what allows for the algorithm based on solving a series of convex optimization problems. Naively introducing a YOY degradation constraint on the problem would break the biconvexity of the problem. In this section, we describe how the non-convexity arises and how the issue is solved mathematically.

Let $d_i$ be the total estimated clear sky energy produced by the system on day $i$ in the data set, and let $T$ be the total number of days in the data set. Enforcing strict yearly periodicity in clear sky daily energy would mean including the following equality constraint to the optimization problem:
\begin{align}
d_{i+365} - d_i = 0 ,\quad\text{for }i=1\ldots T-365,
\end{align}
where each $d_i$ is a decision variable. This is a linear constraint, so if the original subproblems were convex, the new problem with this constraint added would also yield convex subproblems. The most natural way to introduce degradation to the problem (removing strict year-over-year periodicity) would be to introduce an additional decision variable, $\beta$:
\begin{align}
d_{i+365} - d_i = \beta ,\quad\text{for }i=1\ldots T-365,
\end{align}
which allows pairs of days a year apart from each other to differ in daily energy. All pairs of days must differ by the same value, which is found by solving the optimization problem. This relaxation is still a linear constraint. However, YOY degradation is not defined in terms of the difference in energy output but in terms of the percent change in energy output. In other words, we would actually like to introduce the following constraint to the problem: 
\begin{align}
\frac{d_{i+365} - d_i }{d_i} = \beta  ,\quad\text{for }i=1\ldots T-365.
\label{eqn-deg}
\end{align}
This equation contains a ratio of problem variables and is therefore not convex. If we were to naively add this constraint to the \texttt{SCSF} Problem, we would break the biconvex structure. We overcome this difficulty by exploiting the fact that we are using an iterative algorithm to solve the problem. We use bootstrapping to linearize Equation~\ref{eqn-deg} as follows: Let $d^{(j)}\in\mathbf{R}^T$ be the estimate of the clear sky energy output for all days after iteration $j$ of the algorithm, and $d_i^{(j)}\in\mathbf{R}$ is the estimate of day $i$ after iteration $j$. We are using \textit{boostrap} here to mean, ``updating the estimate of a value based on the value found in the previous iteration,'' which is similar to the definition used in Reinforcement Learning~\cite{Sutton2018}. So, the bootstrap-linearized constraint at iteration $j$ becomes
\begin{align}
\frac{d_{i+365}^{(j)} - d_i^{(j)} }{d_i^{(j-1)}} = \beta  ,\quad\text{for }i=1\ldots T-365.
\label{eqn-constraint}
\end{align}
The denominator is the estimate from the previous iteration and is not a problem variable during this iteration. Equation~\ref{eqn-constraint} is a linear constraint in $ d_i^{(j)}$ and $\beta$, so the convexity of the problem is maintained. In the limit $k\rightarrow\infty$, $d_i^{(j)} = d_i^{(j-1)}$, so this $\beta$ converges to the YOY degradation rate in the clear sky response of the system.

\subsection{Tuning parameters}\label{ss:tune-param}

The \texttt{SCSF} problem contains four tuning parameters that control the fit and behavior of the model, as summarized in Table~\ref{tab:tuning-param}. The functional form of \texttt{SCSF} is therefore:
\begin{align}
\hat{p}_{\text{clear sky}} = f(p_{\text{measured}};k,\tau,\mu_L,\mu_R).
\end{align}
It is necessary to correctly set the four model parameters for the function to generate reasonable and useful estimates of clear sky power. The recommended values given in Table~\ref{tab:tuning-param} are based on the parameter sensitivity study presented in Section~\ref{ss:tune-sensitivity}. These values were by performing a grid search over \textit{low}, \textit{medium}, and \textit{high} values, as described in Table~\ref{tab:tuning-study}. The grid search procedure is brute-force optimization heuristic described in~\cite{Bergstra2012}. We select a low number of grid points to start with to reduce computation time. This approach, while not a rigorous optimization, confirmed that the very low sensitivity of the degradation rate estimate to all four of the parameters, and that selected the middle value for each parameter provided a degradation rate estimate near the center of the distribution (see Figure~\ref{fig:grid-dist}). Finally, the parameter $\tau$ was selected for a more detailed study on it's importance. Holding the other parameters constant, the problem was solved over 11 different values of $\tau$ (see Figure~\ref{fig:tau-study}).

\begin{table}[h]
	\centering
	\caption{Summary of \texttt{SCSF} Tuning Parameters}
	\begin{tabular}{|c|l|c|} \hline
		\textbf{Param.} & \textbf{Description} & \textbf{Value} \\
		\hline
		$k$ & Rank of the matrix factorization & $6$ \\
		$\tau$ & Approximate quantile of the data to be fit & $0.85$ \\
		$\mu_L$ & Weight of the smoothing term on the left matrix & $500$ \\
		$\mu_R$ & Weight of the smoothing term on the right matrix & $1000$ \\
		\hline
	\end{tabular}
	\label{tab:tuning-param}
\end{table}

\section{Results}

The algorithm is applied to a data set of 573 residential PV systems. This data was previously compiled by NREL for an analysis \cite{Deceglie2018} that used \texttt{RdTools} to normalize, filter, and analyze the data. For that work, 387 of the 573 systems were selected in a data down-selection process that removed sites with less than two years of data. The NREL study used hourly satellite data for normalization, a major source of uncertainty in the analysis. In Sections~\ref{ss:deg-results} and~\ref{ss:tune-sensitivity}, we reference a number of specific but anonymized systems from this data set to illustrate important results. Section~\ref{ss:fleet-scale} presents the analysis on the entire data set.

\subsection{Fitting the degradation rate}\label{ss:deg-results}

\begin{figure}[t]
	\centering
	\includegraphics[width=\columnwidth]{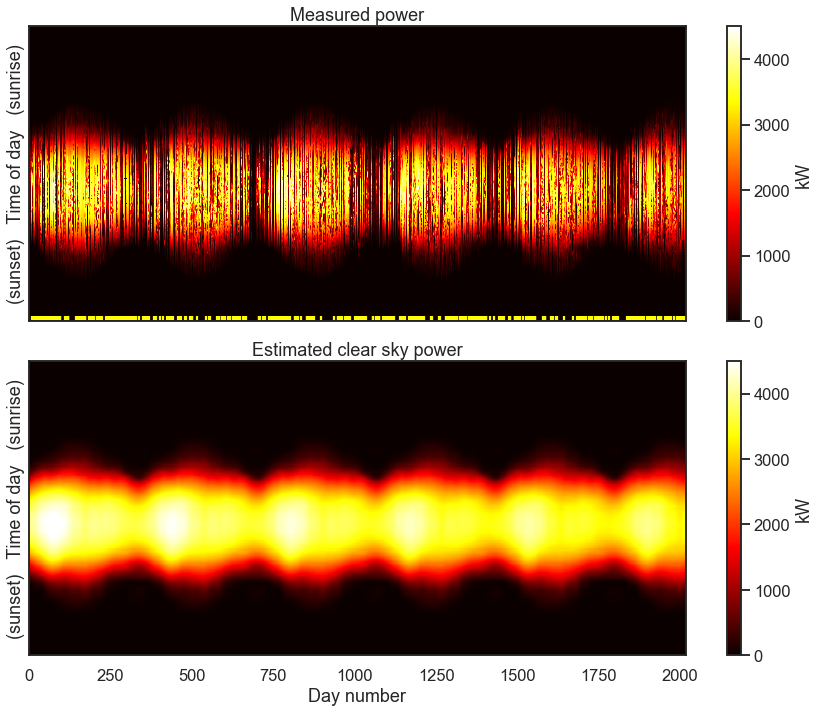}
	\caption{Top: measured time-series data for ES1, viewed as a heatmap. Bottom: estimated clear sky response (with degradation).}
	\label{fig:power_heatmap}
\end{figure}
\begin{figure}[t]
	\centering
	\includegraphics[width=\columnwidth]{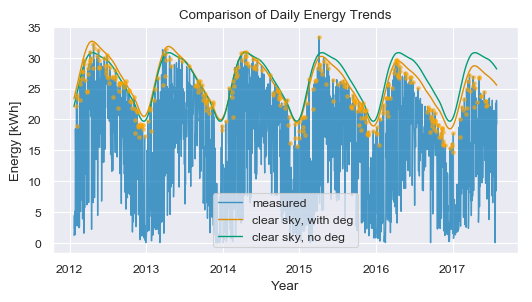} 
	\includegraphics[width=\columnwidth]{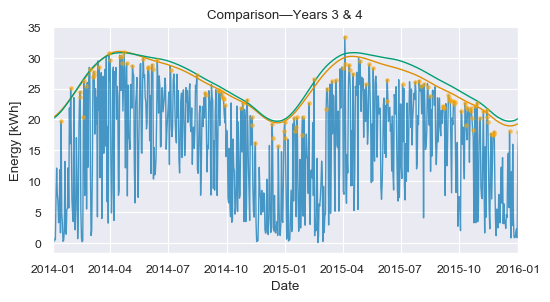}
	\caption{Comparison of summed daily energy, calculated from the observed power signal and from the two estimates of the clear sky response. The same set of parameter values is used for both model implementations. The orange dots indicate roughly clear days in the data set. Note that the summed daily energy in the clear sky models creates an approximate upper envelope fit of the measured data.}
	\label{fig:daily_energy}
\end{figure}
\begin{figure}[t]
	\centering
	\includegraphics[width=\columnwidth]{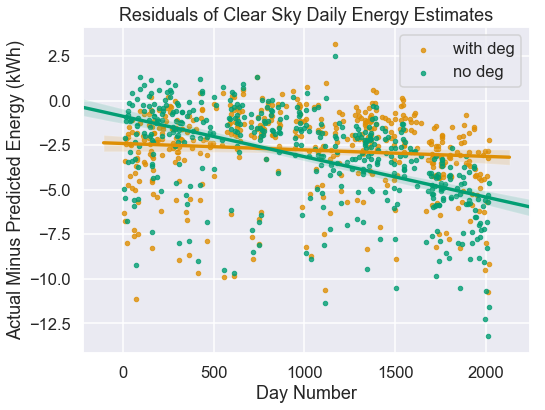}
	\caption{Comparison of the residuals of the \texttt{SCSF} model with a degradation term and without. A simple linear fit is shown for both. The residuals of the fit without a degradation term have a strong dependence on day number, indicating that the model is missing a time-dependent term.}
	\label{fig:residuals}
\end{figure}

We illustrate the need to fit a degradation rate in the \text{SCSF} model by examining Example System 1 (ES1). The power production data and its \texttt{SCSF} baseline are shown in Figure~\ref{fig:power_heatmap}. This example was chosen for its relatively large degradation rate to emphasize the necessity of including degradation in the \texttt{SCSF} model. ES1 has a degradation rate between $-2.4\%$ and $-5.6\%$ according to the \texttt{RdTools} analysis. The \texttt{SCSF} analysis yields a degradation rate of $-2.6\%$. Figure~\ref{fig:daily_energy} shows a comparison of the measured daily energy from ES1 to the estimated clear sky daily energy with and without fitting a degradation rate. The model with the degradation rate is observed to more closely follow the upper envelope of the measured daily energy trend. Figure~\ref{fig:residuals} shows the residuals for the \texttt{SCSF} model with and without a degradation term. Residuals are calcuated as the actually daily energy minus the estimated clear sky daily energy, on approximately clear days only (orange dots in Figure~\ref{fig:daily_energy}). The residuals for the model without a degradation term show a strong dependence on day number, showing that the model is missing a time-dependent factor.

\subsection{Tuning parameter sensitivity}\label{ss:tune-sensitivity}

\begin{figure}[t]
	\centering
	\includegraphics[width=\columnwidth]{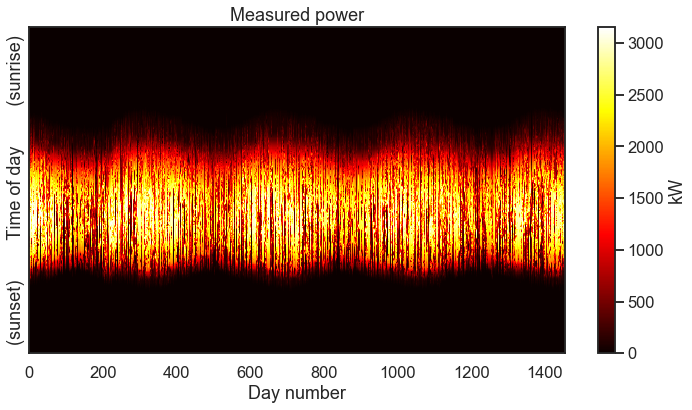}
	\caption{Measured power data for ES2. This system was subjected to a grid search sensitivity study across the four tuning parameters.}
	\label{fig:power-es2}
\end{figure}
\begin{table}[b]
	\centering
	\caption{\texttt{SCSF} Tuning Parameter Grid Search}
	\begin{tabular}{|c|c|c|c|} \hline
		\textbf{Param.} & \textbf{Low Val.} & \textbf{Mid Val.} & \textbf{High Val.} \\
		\hline
		$k$ & $4$ & $6$ & $8$ \\
		$\tau$ & $.8$ & $.85$ & $.9$ \\
		$\mu_L$ & $1e2$ & $5e2$ & $1e3$ \\
		$\mu_R$ & $5e2$ & $1e3$ & $5e3$ \\
		\hline
	\end{tabular}
	\label{tab:tuning-study}
\end{table}

We explore the sensitivity of model fitting to the parameters described in Section~\ref{ss:tune-param} by examining Example System 2 (ES2), whose data is shown in Figure~\ref{fig:power-es2}. This system had a high degree of uncertainty in the \texttt{RdTools} analysis, likely due to the large number of cloudy days. The \texttt{RdTools} degredation estimate for this system is $-2.2\%$, with the 95$^{\text{th}}$ percentile range of $-4.5\%$ to $+0.4\%$.

A grid search was performed over the values summarized in Table~\ref{tab:tuning-study}, resulting in $3^4=81$ \texttt{SCSF} runs on ES2. The average solve time for these runs was 8.7 minutes, for a total linear time of 11.7 hours of computation. This was parallelized over four processors, so the study took about three hours to complete. The high and low values were chosen based on a previous qualitative assessment of the quality of the clear sky estimate over 10 other problem instances (data generated from other systems, randomly selected from the fleet). Outside these high and low values, the clear sky estimate was no longer a reasonable approximation of the possible clear sky response of the system	. 

\begin{figure}[t]
	\centering
	\includegraphics[width=6cm]{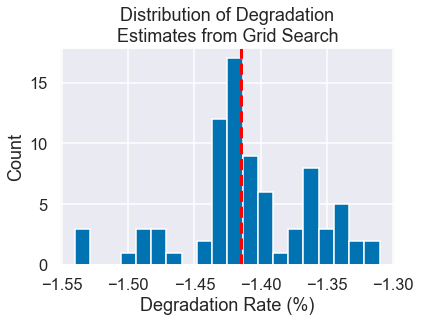}
	\caption{The distribution of degradation rates for ES2 from the grid search study. The red dashed line corresponds to the parameter settings given in Table~\ref{tab:tuning-param}.}
	\label{fig:grid-dist}
\end{figure}

The distribution of the degradation rates obtained from this study is presented in Figure~\ref{fig:grid-dist}. The estimation of degradation is quite stable over these range of values, and the selected values in Table~\ref{tab:tuning-param} correspond to a distinct center peak in the distribution. We interpret these results to mean that the degradation rate for ES2 is $-1.4\%$ with an uncertainty of $\pm 0.1\%$. 

\begin{figure}[t]
	\centering
	\includegraphics[width=\columnwidth]{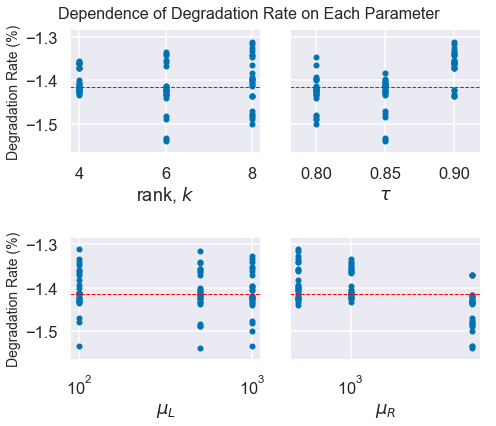}
	\caption{The dependence of the estimated degradation rate on each parameter. The red dashed line corresponds to the degradation rate estimated with the parameter settings given in Table~\ref{tab:tuning-param}.}
	\label{fig:grid-scatter4}
\end{figure}

Figure~\ref{fig:grid-scatter4} shows the dependence of the estimated degradation rate on each tuning parameter. Qualitatively, we found that setting $\mu_L$ and $\mu_R$ too low or too high adversely effected the quality of the clear sky baseline, but did not have a strong effect of the estimated degradation rate. We also found that the rank parameter, $k$, must be sufficiently large such that the model is expressive enough to capture the dynamics in the clear sky signal. Setting $k$ to be too large does not impact the model fitting, but simply adds more variables that must be solved for in the \texttt{SCSF} Problem. Setting $k=6$ was expressive enough for all systems we observed.

\begin{figure}[t]
	\centering
	\includegraphics[width=\columnwidth]{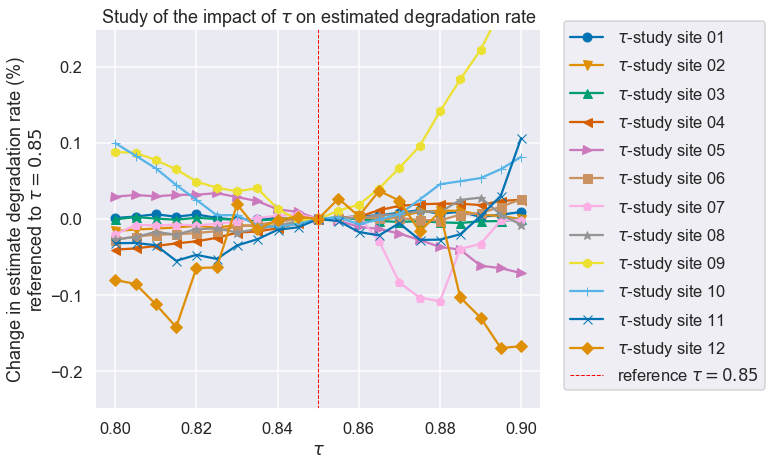}
	\caption{The dependence of the degradation rate on $\tau$ across 13 sites. Because the differences between sites is larger than the variation within a site, the trends are normalized by subtracting the nominal value, i.e. the degradation rate at $\tau=0.85$. All systems show one of the following: monotonic increase, monotonic decrease, a maximum at $\tau\approx 0.85$, or a minimum at $\tau\approx 0.85$.}
	\label{fig:tau-study}
\end{figure}

Finally, the sensitivity on $\tau$ is more significant and more subtle. A more detailed sensitivity study, focusing on $\tau$ was performed for 13 sites from the data set. For these sites, all other parameters were fixed to those given in Table~\ref{tab:tuning-param}, and $\tau$ was varied between $0.8$ and $0.9$ in steps of $0.05$ (21 \texttt{SCSF} runs per site). Over this range of $\tau$, \texttt{SCSF} produces reasonable clear sky baseline estimates, while outside this range the model ceases to yield reasonable estimates. The results of this study are summarize in Figure~\ref{fig:tau-study}. Of the sites included in this study, 6 sites showed variability over this range of less than $0.1\%$, and 12 of the 13 showed variability of less than $0.25\%$. Site $09$ showed the largest variability, with a total range of $0.3\%$.

\begin{figure}[t]
	\centering
	\includegraphics[width=7cm]{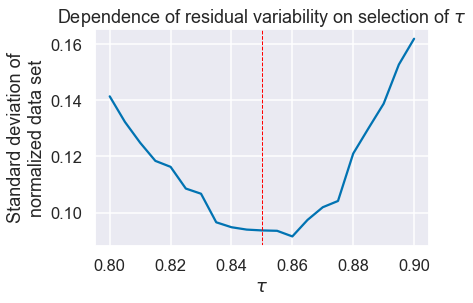}
	\caption{The residual variability in the degradation rate estimates, after norrmalizing to different values of $\tau$. This function has a minimum at $\tau\approx 0.85$.}
	\label{fig:tau-study-residuals}
\end{figure}

Interestingly, all 13 sites exhibit one of four behaviors over this range of $\tau$: (1) monotonic increase, (2) monotonic decrease, (3) a maximum value at around $\tau=0.85$, and (4) a minimal value around $\tau=0.85$. More formally, we find that using $\tau=0.85$ as a normalization point (as in Figure~\ref{fig:tau-study}) approximately minimizes the overall variation in the data. This is illustrated in Figure~\ref{fig:tau-study-residuals} and provides the argument for using $\tau=0.85$ as a default value for fleet-scale analysis.


\subsection{Fleet-scale analysis}\label{ss:fleet-scale}

\begin{figure}[t]
	\centering
	\includegraphics[width=\columnwidth]{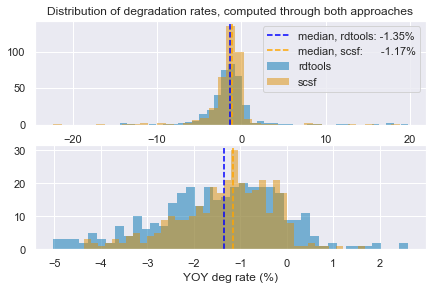}
	\caption{Comparison of YOY system degradation for all systems, as estimated by \texttt{RdTools} and \texttt{SCSF}. The top plot includes all $368$ sites included in both analysis, and the bottom excludes outliers, defined as 1.5 times the interquartile range (332 sites).}
	\label{fig:yoy_distr}
\end{figure}
\begin{figure}[t]
	\centering
	\includegraphics[width=7cm]{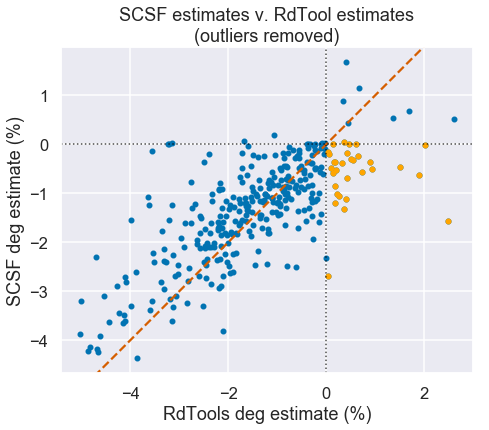}
	\caption{Comparison of \texttt{SCSF} estimates of degradation to \texttt{RdTools} for sites included by both methodologies, with outliers removed (332 sites). The red dashed line is where the two methods agree. Sites that have anomalous, positive estimates from \texttt{RdTools} but negative estimates from \texttt{SCSF} are colored orange.}
	\label{fig:yoy_scatter}
\end{figure}

The \texttt{SCSF} algorithm was applied ``blindly'' to the NREL data set using the parameters summarized in Table \ref{tab:tuning-param}. The complete data set contains data for 573 unique systems from across the continental United States. The \texttt{RdTools} approach rejected 186 sites, while the \texttt{SCSF} methodology rejected 22 sites. Only three sites were rejected by both methods; all other rejected sites were unique to that method. 368 sites were included in both analyses. Table \ref{tab:fleet-join} summarizes these results.

\begin{table}[b]
	\centering
	\caption{Summary of site selection for both methods}
	\begin{tabular}{|c|c|c|c|} \hline
		 &  & & \textbf{Unique} \\
		 & \textbf{Included Sites}& \textbf{Excluded Sites} & \textbf{Excluded Sites}  \\
		\hline
		\texttt{SCSF}& $551$ & $22$ & $19$  \\
		\texttt{RdTools} & $387$ & $186$ & $183$\\
		\hline
	\end{tabular}
	\label{tab:fleet-join}
\end{table}

Figure~\ref{fig:yoy_distr} shows the distribution of YOY degradation rates across all systems in the NREL data set from both the \texttt{RdTools} approach and the \texttt{SCSF} approach. The median values of two methods are in very close agreement, validating that the two methods are measuring the same fundamental quantity, in expectation. After excluding outliers (bottom plot), the standard deviation of the \texttt{SCSF} distribution is $1.0\%$ versus $1.4\%$ for \texttt{RdTools}, indicating a reduction in uncertainty of the fleet-scale analysis. An outlier here is defined based Tukey's definition~\cite{tukey1977}: If the quartiles of the data set are $Q_1$ for the first quartile and $Q_3$ for the third quartile, then outliers are points that fall outside the following interval:
\begin{align*}
\left[Q_1-1.5\cdot(Q_3 - Q_1),\, Q_3+1.5\cdot(Q_3-Q_1)\right]
\end{align*}

Figure \ref{fig:yoy_scatter} compares the \texttt{SCSF} estimates to the \texttt{RdTools} estimates. Outliers have been removed, as defined above. On a site-by-site basis, the two methods can differ on the order of $2\%$.  Each \texttt{RdTools} estimate should be considered the P50 degradation rate in the context of an associated confidence interval~\cite{Deceglie2018}, and $85.6\%$ of the sites have an \texttt{SCSF} estimate that is within the confidence bounds for the \texttt{RdTools} approach. There is a significant population of sites corresponding to positive \texttt{RdTools} estimates and negative \texttt{SCSF} estimates, colored in orange. Note that there is not a corresponding population in quadrant 2 (positive \texttt{SCSF}, negative \texttt{RdTools}). The \texttt{SCSF} analysis resulted in positive degradation rates for 20 sites, while the \texttt{RdTools} analysis found positive degradation for 50 sites. For crystalline silicon based systems, such as those considered here, energy output of a system is generally not expected to increase over time, so positive ``degradation'' rates typically indicate an error with the analysis or a lack of robustness to data anomalies.

\begin{figure}[t]
	\centering
	\includegraphics[width=\columnwidth]{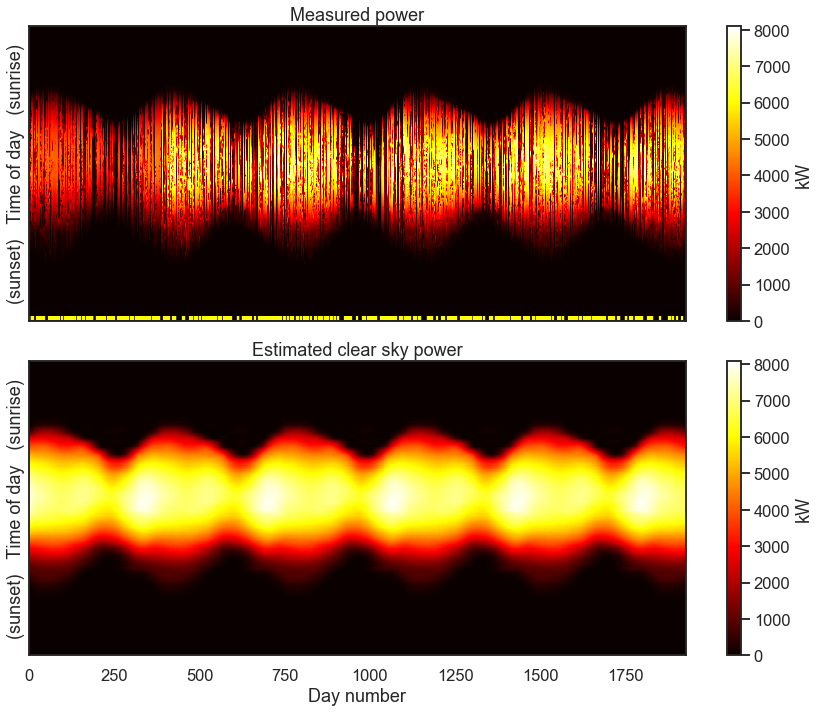}
	\caption{Top: measured time-series data for ES3, viewed as a heatmap (note the abnormal, lower power output in the first year). Bottom: estimated clear sky response (with degradation). Note how the clear sky signal robustly ignores the reduced power output in year 1.}
	\label{fig:power_heatmap_ES3}
\end{figure}

\begin{figure}
	\centering
	\includegraphics[width=\columnwidth]{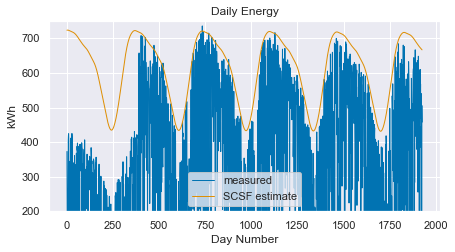}
	\caption{Comparison of summed daily energy, calculated from the observed power signal and from the \texttt{SCSF} clear sky estimate. Note that the algorithm is completely robust to the data anomaly in the first year.}
	\label{fig:daily_energy_B}
\end{figure}

An interesting case study for contrasting the \texttt{SCSF} and \texttt{RdTools} approaches is Example System 3 (ES3). The measured power and fit clear sky model for this system are shown in Figure~\ref{fig:power_heatmap_ES3}. This system shows a different apparent capacity in the first year than the subsequent four years. This type of data anomaly is quite common and can be caused by someone updating a scale factor in the data logger program or by simply installing more panels (the cause in this case is unknown).  Figure~\ref{fig:daily_energy_B} illustrates the robustness of the \texttt{SCSF} method to this data anomaly. The clear sky estimator completely ignores the data in the first year, focusing the fit on the later data. Notably, the \texttt{RdTools} approach estimated the degradation rate for this system to be $+11.8\%$, whereas \texttt{SCSF} estimate is approximately zero which appears to be a more reasonable estimate of YOY degradation for this system.


\section{Conclusion}

We present an extension to Statistical Clear Sky Fitting \cite{Meyers2018} that allows for efficient estimation of system YOY degradation rate. This approach to estimating degradation requires no model of the site nor any estimates of irradiance or weather data. For this reason, it lends itself naturally to fleet-scale analysis of heterogeneous PV systems, where such supplementary data may be missing or incorrect. 

We show that fitting a periodic data model to measured PV power data requires the inclusion of a degradation term, and we demonstrate the sensitivity of that degradation term to model tuning parameters. The \texttt{SCSF} approach to degradation estimation is at least as accurate, on average, as the \texttt{RdTools} approach. In addition, the analysis of this fleet of PV systems suggests that the \texttt{SCSF} approach yields improved accuracy by reducing the variance of the degradation rates across all systems and by reducing the number of systems with non-negative degradation rates. Both these effects relate to improved robustness of \texttt{SCSF} to data anomalies relative to \texttt{RdTools}.

The ability to perform model-less estimation of bulk PV system degradation rates presents a large opportunity for evaluating the value proposition of fleet-scale collections of distributed rooftop systems. This algorithm enables degradation analysis for a broad set of systems where data to supplement the system power signals is unavailable, unreliable, or expensive to procure. In addition, the \texttt{SCSF} method requires no engineering time to model the system, merge PV power and weather data sets, or filter data. The approach is naturally robust to missing and bad power data, handling common anomalies such as missing data and changes in apparent system capacity.

The \texttt{SCSF} algorithm provides not only an estimate of system degradation but also an estimate of the clear sky response of the system, which can be thought of as an optimal baseline capturing the cyclostationarity of the observed time-series power signal. In addition, the biconvex structure of the \texttt{SCSF} approach lends itself to efficient computation and scalability. \text{SCSF} represents a fundamentally new way of modeling and understanding PV power data sets, exploiting signal processing techniques and periodicity rather than physical models to understand system behavior. The software implementation of \texttt{SCSF} is available at \url{https://github.com/slacgismo/StatisticalClearSky}. Data cleaning and preprocessing was automated with software available at \url{https://github.com/slacgismo/solar-data-tools}.

\section*{Acknowledgments}

Thank you to Laura Schelhas and Elpiniki Apostolaki Iosifidou, who collaborate on the PVInsight project, and to the rest of the GISMo team at SLAC National Accelorator Laboratory. Additional thanks to Prof. Stephen Boyd at Stanford University for the support and inspiration on this work. This material is based upon work supported by the U.S. Department of Energy's Office of Energy Efficiency and Renewable Energy (EERE) under Solar Energy Technologies Office (SETO) Agreement Numbers 34911 (``PVInsight''), 30311, and 34348. This work was coauthored by Alliance for Sustainable Energy, LLC, the manager and operator of the National Renewable Energy Laboratory for the US Department of Energy (DOE) under Contract No. DE‐AC36‐08GO28308. The views expressed in the article do not necessarily represent the views of the DOE or the US Government.  The US Government retains and the publisher, by accepting the article for publication, acknowledges that the US Government retains a nonexclusive, paid‐up, irrevocable, worldwide license to publish or reproduce the published form of this work, or allow others to do so, for US Government purposes.	

\small
\bibliographystyle{IEEEtran}
\bibliography{PVSC2019}

\begin{thebibliography}{10}
\providecommand{\url}[1]{#1}
\csname url@samestyle\endcsname
\providecommand{\newblock}{\relax}
\providecommand{\bibinfo}[2]{#2}
\providecommand{\BIBentrySTDinterwordspacing}{\spaceskip=0pt\relax}
\providecommand{\BIBentryALTinterwordstretchfactor}{4}
\providecommand{\BIBentryALTinterwordspacing}{\spaceskip=\fontdimen2\font plus
\BIBentryALTinterwordstretchfactor\fontdimen3\font minus
  \fontdimen4\font\relax}
\providecommand{\BIBforeignlanguage}[2]{{%
\expandafter\ifx\csname l@#1\endcsname\relax
\typeout{** WARNING: IEEEtran.bst: No hyphenation pattern has been}%
\typeout{** loaded for the language `#1'. Using the pattern for}%
\typeout{** the default language instead.}%
\else
\language=\csname l@#1\endcsname
\fi
#2}}
\providecommand{\BIBdecl}{\relax}
\BIBdecl

\bibitem{Phinikarides2014}
\BIBentryALTinterwordspacing
A.~Phinikarides, N.~Kindyni, G.~Makrides, and G.~E. Georghiou, ``{Review of
  photovoltaic degradation rate methodologies},'' \emph{Renewable and
  Sustainable Energy Reviews}, vol.~40, pp. 143--152, 2014. [Online].
  Available: \url{http://dx.doi.org/10.1016/j.rser.2014.07.155}
\BIBentrySTDinterwordspacing

\bibitem{Hasselbrink2013}
E.~Hasselbrink, M.~Anderson, Z.~Defreitas, M.~Mikofski, Y.~C. Shen,
  S.~Caldwell, A.~Terao, D.~Kavulak, Z.~Campeau, and D.~Degraaff, ``{Validation
  of the PVLife model using 3 million module-years of live site data},''
  \emph{Conference Record of the IEEE Photovoltaic Specialists Conference}, pp.
  7--12, 2013.

\bibitem{Jordan2018}
D.~C. Jordan, C.~Deline, S.~R. Kurtz, G.~M. Kimball, and M.~Anderson, ``{Robust
  PV Degradation Methodology and Application},'' \emph{IEEE Journal of
  Photovoltaics}, vol.~8, no.~2, pp. 525--531, 2018.

\bibitem{rdtools}
\BIBentryALTinterwordspacing
``{RdTools v1.2.2}.'' [Online]. Available:
  \url{https://github.com/NREL/rdtools}
\BIBentrySTDinterwordspacing

\bibitem{King2004}
\BIBentryALTinterwordspacing
D.~L. King, W.~E. Boyson, and J.~A. Kratochvil, ``{Photovoltaic array
  performance model},'' \emph{Sandia Report No. 2004-3535}, vol.~8, no.
  December, pp. 1--19, 2004. [Online]. Available:
  \url{http://prod.sandia.gov/techlib/access-control.cgi/2004/043535.pdf}
\BIBentrySTDinterwordspacing

\bibitem{Meyers2018}
\BIBentryALTinterwordspacing
B.~Meyers, M.~Tabone, and E.~C. Kara, ``{Statistical Clear Sky Fitting
  Algorithm},'' \emph{Conference Record of the IEEE Photovoltaic Specialists
  Conference}, pp. 1--6, 2018. [Online]. Available:
  \url{http://www.wcpec7.org/eWCPEC/manuscripts/MeWCPEC930_0521112519.pdf}
\BIBentrySTDinterwordspacing

\bibitem{Udell2016}
\BIBentryALTinterwordspacing
M.~Udell, C.~Horn, R.~Zadeh, and S.~Boyd, ``{Generalized Low Rank Models},''
  \emph{Foundations and Trends in Machine Learning}, vol.~9, no.~1, pp. 1--118,
  2016. [Online]. Available:
  \url{https://web.stanford.edu/~boyd/papers/glrm.html}
\BIBentrySTDinterwordspacing

\bibitem{Deceglie2018}
M.~Deceglie, D.~C.~Jordan, A.~Nag, A.~Shinn, and C.~Deline, ``Fleet-scale
  energy-yield degradation analysis applied to hundreds of residential and
  nonresidential photovoltaic systems,'' \emph{IEEE Journal of Photovoltaics},
  vol.~PP, pp. 1--7, 01 2019.

\bibitem{Brillinger1981}
\BIBentryALTinterwordspacing
D.~R. Brillinger, \emph{{Time Series: Data Analysis and Theory}}, ser. Classics
  in Applied Mathematics.\hskip 1em plus 0.5em minus 0.4em\relax Philadelphia,
  PA: Society for Industrial and Applied Mathematics (SIAM), 1981, p.~1.
  [Online]. Available: \url{https://books.google.com/books?id=3DFJfgEW94gC}
\BIBentrySTDinterwordspacing

\bibitem{Hastie2009}
T.~Hastie, R.~Tibshirani, and J.~Friedman, \emph{The Elements of Statistical
  Learning}, ser. Springer Series in Statistics.\hskip 1em plus 0.5em minus
  0.4em\relax New York, NY, USA: Springer New York Inc., 2009, pp. 33--36.

\bibitem{Davies2008}
\BIBentryALTinterwordspacing
P.~Davies and M.~Meise, ``Approximating data with weighted smoothing splines,''
  \emph{Journal of Nonparametric Statistics}, vol.~20, no.~3, pp. 207--228,
  2008. [Online]. Available: \url{https://doi.org/10.1080/10485250801948625}
\BIBentrySTDinterwordspacing

\bibitem{Muller1987}
\BIBentryALTinterwordspacing
H.-G. Müller, ``Weighted local regression and kernel methods for nonparametric
  curve fitting,'' \emph{Journal of the American Statistical Association},
  vol.~82, no. 397, pp. 231--238, 1987. [Online]. Available:
  \url{https://doi.org/10.1080/01621459.1987.10478425}
\BIBentrySTDinterwordspacing

\bibitem{Pearson1901}
K.~Pearson, ``{On lines and planes of closest fit to systems of points in
  space},'' \emph{The London, Edinburgh, and Dublin Philosophical Magazine and
  Journal of Science}, vol.~2, no.~11, pp. 559--572, 1901.

\bibitem{Eckart1936}
C.~Eckart and G.~Young, ``{The approximation of one matrix by another of lower
  rank},'' \emph{Psychometrika}, vol.~1, no.~3, pp. 211--218, 1936.

\bibitem{Tosic2011}
I.~Tosic and P.~Frossard, ``{Dictionary Learning},'' \emph{IEEE Signal
  Processing Magazine}, vol.~28, no.~2, pp. 27--38, 2011.

\bibitem{Reda2004}
\BIBentryALTinterwordspacing
I.~Reda and A.~Andreas, ``Solar position algorithm for solar radiation
  applications,'' \emph{Solar Energy}, vol.~76, no.~5, pp. 577 -- 589, 2004.
  [Online]. Available:
  \url{http://www.sciencedirect.com/science/article/pii/S0038092X0300450X}
\BIBentrySTDinterwordspacing

\bibitem{Reda2007}
\BIBentryALTinterwordspacing
------, ``Corrigendum to “solar position algorithm for solar radiation
  applications” [solar energy 76 (2004) 577–589],'' \emph{Solar Energy},
  vol.~81, no.~6, p. 838, 2007. [Online]. Available:
  \url{http://www.sciencedirect.com/science/article/pii/S0038092X07000059}
\BIBentrySTDinterwordspacing

\bibitem{INEICHEN2008}
\BIBentryALTinterwordspacing
P.~Ineichen, ``A broadband simplified version of the solis clear sky model,''
  \emph{Solar Energy}, vol.~82, no.~8, pp. 758 -- 762, 2008. [Online].
  Available:
  \url{http://www.sciencedirect.com/science/article/pii/S0038092X08000406}
\BIBentrySTDinterwordspacing

\bibitem{INEICHEN2016}
\BIBentryALTinterwordspacing
------, ``Validation of models that estimate the clear sky global and beam
  solar irradiance,'' \emph{Solar Energy}, vol. 132, pp. 332 -- 344, 2016.
  [Online]. Available:
  \url{http://www.sciencedirect.com/science/article/pii/S0038092X16002048}
\BIBentrySTDinterwordspacing

\bibitem{Reno2016}
\BIBentryALTinterwordspacing
M.~J. Reno and C.~W. Hansen, ``{Identification of periods of clear sky
  irradiance in time series of GHI measurements},'' \emph{Renewable Energy},
  vol.~90, pp. 520--531, 2016. [Online]. Available:
  \url{http://dx.doi.org/10.1016/j.renene.2015.12.031}
\BIBentrySTDinterwordspacing

\bibitem{Curran2019}
A.~J. Curran, C.~B. Jones, S.~Lindig, J.~Stein, D.~Moser, and R.~H. French,
  ``{Performance Loss Rate Consistency and Uncertainty Across Multiple Methods
  and Filtering Criteria},'' \emph{Conference Record of the IEEE Photovoltaic
  Specialists Conference}, 2019.

\bibitem{Gorski2007}
J.~Gorski, F.~Pfeuffer, and K.~Klamroth, ``{Biconvex sets and optimization with
  biconvex functions: A survey and extensions},'' \emph{Mathematical Methods of
  Operations Research}, vol.~66, no.~3, pp. 373--407, 2007.

\bibitem{Sutton2018}
\BIBentryALTinterwordspacing
R.~S. Sutton and A.~G. Barto, \emph{{Reinforcement Learing - An
  Introduction}}.\hskip 1em plus 0.5em minus 0.4em\relax Cambridge, MA: MIT
  Press, 2018, p.~89. [Online]. Available:
  \url{http://incompleteideas.net/book/the-book-2nd.html}
\BIBentrySTDinterwordspacing

\bibitem{Bergstra2012}
J.~Bergstra and Y.~Bengio, ``{Random Search for Hyper-Parameter
  Optimization},'' \emph{Journal of Machine Learning Research}, vol.~13, pp.
  281--305, 2012.

\bibitem{tukey1977}
J.~W. Tukey, \emph{Exploratory data analysis}, ser. Addison-Wesley series in
  behavioral sciences.\hskip 1em plus 0.5em minus 0.4em\relax Reading, MA:
  Addison-Wesley Pub. Co., 1977.

\end{thebibliography}

\end{document}